\documentclass[
 aps, pra,
 amsmath,amssymb,
 11pt,
 final,
tightenlines,
 twoside,
 twocolumn,
 nofloats,
nofootinbib,
 superscriptaddress,
showkeys,
showkeywords,
 ]
{revtex4-2}

\renewcommand{\thetable}{\arabic{table}}

\usepackage[T2A]{fontenc}
\usepackage[utf8]{inputenc}
\usepackage[english]{babel}
\usepackage{graphicx}% Include figure files
\usepackage{dcolumn}% Align table columns on decimal point
\usepackage{bm}% bold math

\usepackage{xcolor}
\usepackage{soul}
\usepackage{changes}
\usepackage{hyperref}

\setcitestyle{authoryear,round}
\def\kms{km\,s$^{-1}${}}
\def\mcm{$\mu$m{}}
\def\IRAS{IRAS~21204+4913{}}
\def\Av{A_{\text V}{}}
\def\degr{\hbox{$^\circ$}}
\def\arcsec{\hbox{$^{\prime\prime}$}}
\def\fm{\hbox{$.\!\!^{\rm m}$}}

\def\farcs{\hbox{$.\!\!^{\prime\prime}$}}
\def\farcm{\hbox{$.\mkern-4mu^\prime$}}
\DeclareRobustCommand{\ion}[2]{\textup{#1\,\textsc{\lowercase{#2}}}}

\def\squareforqed{\hbox{\rlap{$\sqcap$}$\sqcup$}}

\def\sq{\ifmmode\squareforqed\else{\unskip\nobreak\hfil
\penalty50\hskip1em\null\nobreak\hfil\squareforqed
\parfillskip=0pt\finalhyphendemerits=0\endgraf}\fi}

\def\degr{\hbox{$^\circ$}}

\def\arcsec{\hbox{$^{\prime\prime}$}}

\def\utw{\smash{\rlap{\lower5pt\hbox{$\sim$}}}}

\def\udtw{\smash{\rlap{\lower6pt\hbox{$\approx$}}}}

\def\fm{\hbox{$\,.\!\!^{\rm m}$}}

\def\farcm{\hbox{$\,.\mkern-4mu^\prime$}}

\def\farcs{\hbox{$\,.\!\!^{\prime\prime}$}}

\def\diameter{{\ifmmode\mathchoice
{\ooalign{\hfil\hbox{$\displaystyle/$}\hfil\crcr
{\hbox{$\displaystyle\mathchar"20D$}}}}
{\ooalign{\hfil\hbox{$\textstyle/$}\hfil\crcr
{\hbox{$\textstyle\mathchar"20D$}}}}
{\ooalign{\hfil\hbox{$\scriptstyle/$}\hfil\crcr
{\hbox{$\scriptstyle\mathchar"20D$}}}}
{\ooalign{\hfil\hbox{$\scriptscriptstyle/$}\hfil\crcr
{\hbox{$\scriptscriptstyle\mathchar"20D$}}}}
\else{\ooalign{\hfil/\hfil\crcr\mathhexbox20D}}%
\fi}}

\begin{document}


%\ydk{}
%\titlerunning{}
%\authorrunning{}
%\toctitle{}
%\tocauthor{}
{\it Accepted by Astronomy Letters}

\title{THE ERUPTIVE YOUNG STAR \IRAS}

\author{\firstname{M.~A.}~\surname{Burlak}}
\author{\firstname{A.~V.}~\surname{Dodin}}
\author{\firstname{A.~V.}~\surname{Zharova}}
\author{\firstname{S.~G.}~\surname{Zheltoukhov}}
\author{\firstname{N.~P.}~\surname{Ikonnikova}}
\author{\firstname{S.~A.}~\surname{Lamzin}}
\email{lamzin@sai.msu.ru}
\author{\firstname{B.~S.}~\surname{Safonov}}
\author{\firstname{I.~A.}~\surname{Strakhov}}
\author{\firstname{A.~A.}~\surname{Tatarnikov}}
\author{\firstname{A.~M.}~\surname{Tatarnikov}}
\affiliation{Sternberg Astronomical Institute, M.V. Lomonosov Moscow State University, Moscow, 119234 Russia\\}

     %%%%%%%%%%%%%%%%%%%%%%%%%%%%%%%%%%%%%%%%%%%%%%

\begin{abstract}
The results of photometric, polarimetric, and spectroscopic observations are presented for the young star IRAS 21204+4913, whose visible brightness has increased by $\approx 5^{\rm m}$ since October 2025. The star's absorption spectrum in the $0.36 - 0.75~\mu$m range resembles those of A--F giants and supergiants, but it also exhibits molecular TiO bands. The brightening was accompanied by a significant increase in the degree of polarization of the stellar radiation (to $\approx 16\,\%$ in the $I$-band), likely due to 
scattering by dust in an expanding circumstellar shell. The P~Cygni profile of the H$\alpha$ line implies a dusty wind velocity of $\approx ~300$ {\kms}.
We believe that the outburst of IRAS 21204+4913 is caused by an increase in the accretion rate of protoplanetary disk's matter onto the young star with
a mass of $\lesssim 0.5$~M$_\odot$ to $\gtrsim 3\times 10^{-5}$~M$_\odot$\,yr$^{-1}$. Furthermore, IRAS 21204+4913 displays several unusual features: 
the dependence of the width and radial velocity of absorption lines on the excitation potential, emission in the TiO molecular bands, and a comparably 
bright outburst that occured in 1948. Several T Tauri stars and a group of Herbig-Haro objects are found in the vicinity of the star.
\end{abstract}
         %%%%%%%%%%%%%%%%%%%%%%%%%%%%%%%%%%%%%%%%%%%%
\maketitle

\section{Introduction}
\label{sect:introduct}
During the transformation of a protostellar cloud into 
a T~Tauri star -- a process lasting $\sim0.5$~Myr -- 
some young stellar objects (YSOs) undergo dramatic 
brightening events that can persist from several months 
to several decades \citep{Herbig-1977}. 
It is generally accepted that outbursts of this kind are
caused by a strong increase in the accretion rate 
$\dot M_{\text ac}$ of protoplanetary disk matter onto 
the forming star \citep{Hartmann-Keyon-1996, Hartmann-2016, Audard-2014}. In extreme cases, $\dot M_{\text ac}$ can 
exceed $10^{-4}$~M$_\odot$\,yr$^{-1}$, which is thousands 
of times more than typical $\dot M_{\text ac}$ observed in classical T~Tauri stars (CTTSs). If such events recur frequently during star formation, they may significantly influence the final stellar mass and other fundamental parameters of the star -- see \citet{Baraffe-2017, Fischer-2023} and references therein. 
A strong, albeit relatively short-term, increase in the luminosity of YSOs undoubtedly influences the evolution of their protoplanetary disks, and therefore the formation of planetary systems -- see, e.g., \citet{Stamatellos-2012, Topchieva-2025}.

Various mechanisms have been proposed to explain the origin of powerful episodic accretion. Without aiming for completeness (see e.g. \citealt{Fischer-2023}), we mention only a few: thermal, gravitational, and/or magnetorotational disk instabilities \citep{Zhu-2010, Bae-2014, Meyer-2017, Kadam-2020}; interactions between the disk and a planet or stellar companion \citep{Borchert-2022, Nayakshin-2023}; an infall of gas
cloud onto the disk \citep{Demidova-2023}.

Depending on the outburst amplitude, duration, and spectral characteristics, \citet{Herbig-1977} divided eruptive YSOs into two classes: FU~Orionis-type objects -- FUors\footnote{The term {\it fuor} was introduced by V.~A.~Ambartsumyan (\citeyear{Ambartsumyan-1971}).} 
and EX~Lupi-type objects -- EXors \citep{Herbig-1989}. 
Herbig classified FUors as objects whose visible brightness increased by $4^{\text m} - 6^{\text m}$ over several months and then gradually decreased over several decades, and whose spectra were almost entirely absorption-dominated throughout the outburst. 
EXors, in contrast, exhibit recurrent outbursts -- sometimes reaching FUor-like amplitudes but lasting only weeks to a year -- during which their spectra are dominated by emission lines.

Over the past decade, the advent of numerous monitoring systems has led to the discovery of a large number of eruptive YSOs. A team of researchers \citet{Contreras-2025} compiled the OYCAT (Outbursting YSOs Catalogue)\footnote{\url{http://starformation.synology.me:5002/OYCAT/main.html}}, which currently contains basic information on 174 objects. According to the established criteria \citep[][table~1]{Contreras-2025}, the objects in the catalogue are divided into several groups based on the photometric and spectral characteristics of their outbursts. Without going into detail, we note that the authors confidently classified as FUors and EXors fewer than 40\,\%{} of all objects, whereas nearly half of the YSOs exhibited outburst parameters intermediate between those of FUors and EXors.

The diversity of observed manifestations of episodic accretion can be attributed either to different underlying physical mechanisms or to the combined action of multiple mechanisms -- see, for example, \citet{Skliarevskii-2023}. Acquiring the most comprehensive possible information on each eruptive YSO will help elucidate the general physics of the phenomenon. In this context, we became interested in the report by S. Kaneko\footnote{\url{https://www.astroarts.co.jp/photo-gallery/photo/129147}}, who discovered on November 25, 2025, the optical transient TCP~J21220926+4926242 (hereafter J2122), whose brightness in the optical band approximately doubled over the following week.

Subsequently, it became clear that the outburst occurred no later than October 25, 2025, when the transient was first detected by the ASAS-SN survey \citep{Shappee-2014} at a magnitude of $g=16\fm 9$ \citep{Kochkina-2025}. By late November, when the brightness of J2122 exceeded $V\approx 12\fm 4$, \citet{Kochkina-2025} carried out the first spectroscopic observations of the transient in the wavelength range of $0.36-0.72$~{\mcm} and found that its spectrum resembled the absorption-dominated spectra typical of FUor-like objects: the only emission feature detected in the spectrum of J2122 was the [\ion{S}{II}]~6731~\AA{} 
line.

\citet{Grosso-2025} identified the transient J2122 with 
the IRAS source 21204+4913 \citep{Neugebauer-1984}, which, according to \citet{Marton-2016}, is a class I/II YSO -- 
i.e., it is in the transitional evolutionary stage between a protostar and a pre-main-sequence star \citep{Adams-Lada-Shu-1987, Lada-2005}. {\IRAS} lies on the southeastern edge of the relatively compact (18 square arcminutes) dark nebula 
D\,2944 \citep{Dobashi-2011}. \citet{Grosso-2025} 
determined a distance to the source of $\approx 500$~pc, and \citet{Stecklum-2025} reported a line-of-sight extinction of $A_{\rm V} \approx 8^{\text m}.$

We began observing {\IRAS} in early December 2025, and in this paper we present the results of our research. First, we describe our observational data and the results derived from them; then, we examine what our data reveal about the nature of {\IRAS}. The conclusions and main findings of this work are summarized in the Conclusion.
 
			%%%%%%%%% End "Introduction" section %%%%%%%%%%

	        %%%%%%%%%%%  Sect: Observations %%%%%%%%%%
%
\section{Observations}\label{sect:observation}

To construct the historical light curve of {\IRAS}, we selected approximately 100 photographic plates from the collection of the Sternberg Astronomical Institute of Lomonosov Moscow State University (SAI MSU) that included the star. These plates were obtained between May 10, 1899, and November 7, 1975, in a photometric system close to the Johnson $B$ band. The limiting magnitude of the plates varied from $B\approx 14^{\text m}$ to $18^{\text m}$, depending on the instrument used and the observing conditions.

The observations presented below were carried out in 2025--2026 with the instruments of the Caucasian Mountain Observatory (CMO) of SAI MSU \citep{Shatsky-2020}. Optical photometry of {\IRAS} was obtained with the 60-cm RC600 telescope equipped with a CCD camera and a set of $UBVR_{\text c}I_{\text c}$ filters in the Bessell-Cousins photometric system \citep{Berdnikov2020}. The magnitudes of the comparison stars were calibrated against the standard field GD391 \citep{Clem-Landolt-2016}. The results of the optical photometry are presented in Table~\ref{tab:Optph}.

Photometric observations of the star in the near-infrared (NIR) were carried out in the $YJHK$ bands of the MKO-NIR photometric system using the ASTRONIRCAM infrared camera-spectrograph \citep{Nadjip-17} mounted on the 2.5-m telescope. A detailed description of the observational procedure and data reduction can be found in \citet{Tatarnikov-2023}. The following stars from the 2MASS catalogue were used as photometric standards: 2MASS J21222481+4927597, 2MASS J21220594+4925335, and 2MASS J21222134+4925561. Their magnitudes were transformed into the MKO-NIR system using the color-term equations provided in \citet{Tat_Tat-2023}. Additionally, we observed the source in the $L^\prime$ $(\lambda_{\text{c}} = 3.8$~\mcm) and $M^\prime$ $(\lambda_{\text{c}} = 4.7$~\mcm) bands -- close to the corresponding MKO-NIR bands -- using the LMP camera \citep{Zheltoukhov-2024}. Photometric reference stars from the list of \citet{Shenavrin-2011}, positioned at the same airmass as the target during the observations, were employed as comparison stars. The results of our NIR photometry are presented in Table~\ref{tab:IRph}.

Polarimetric observations in the $BVR_{\text c}I_{\text c}$ bands 
were carried out with the 2.5-m telescope using the speckle polarimeter \citep{Safonov-2017}. The instrument is a dual-beam polarimeter equipped with a rotating half-wave plate as the modulator and a polarizing beamsplitting cube as the beamsplitter. The primary detector is a fast, low-noise CMOS camera (Hamamatsu ORCA-Quest).
For each photometric band, we obtained a series of 4000 frames, each with an exposure time of 60~ms. During the acquisition, the half-wave plate rotated continuously at an angular velocity of 60$^{\circ}$\,s$^{-1}$.

The acquired images were corrected for bias offset and demodulated using the known position angle of the half-wave plate for each frame. Aperture photometry was then performed on the resulting demodulated series, taking into account the possible contribution from polarized background emission. The resulting flux measurements were used to estimate the Stokes parameters in the instrumental reference frame. Transformation to the equatorial reference frame and correction for instrumental polarization of the telescope (observations were carried out in the Nasmyth-2 focus) were performed following the procedure described in \citet{Safonov-2017}. The results are presented in Table~\ref{tab:polariz} and Fig.~\ref{fig:PvsIWL}.

Low-resolution spectra of {\IRAS} were obtained with the Transient Double-beam Spectrograph (TDS). A detailed description of the instrument and the data reduction procedure can be found in \citet{Potanin-2020}. Here we only note that the spectral resolving power $R=\lambda/\Delta \lambda$ of the TDS is $\approx1300$ in the red channel ($0.56-0.74$~\mcm) and $\approx 2400$ in the blue channel ($0.36-0.56$~\mcm) \citep{Belinski-2023}.

High-resolution spectra (HRS, $R\sim20\, 000$ over the wavelength range $4050-8300$~{\AA}) were obtained with the ``Raduga'' spectrograph -- an \'{e}chelle fiber-fed spectrograph with a fiber diameter of $2\farcs5$. Data reduction was performed in the standard manner. Wavelength calibration was carried out using a hollow-cathode lamp (HCL). Spectral extraction followed the method described in \citet{Bolton-2010}, employing HCL emission lines to construct a two-dimensional instrumental line-spread function. The high spatial resolution of the detector ($9566\times6388$~pixels) enables sufficiently detailed modeling of the line profile to solve the inverse problem of spectrum extraction. Wavelength calibration frames were acquired immediately before and after each science exposure, and the dispersion solution was interpolated to the mid-exposure time of the science frame. Stability tests using radial velocity standard stars (HIP\,43726 and HIP\,17378) showed that the local calibration accuracy is approximately $\sigma=0.4$~\kms{} -- this represents the scatter of individual spectral lines in the Raduga spectra relative to a reference spectrum obtained with the ESPRESSO spectrograph \citep{Pepe-2021} at $R \sim 190\,000$. The random scatter in the mean radial velocities of the standard stars observed on different nights is $\sigma \approx 0.03$~\kms{}.

All wavelengths and observation times for both spectrographs have been corrected to the barycenter of the Solar System. The log of spectral observations is presented in Table~\ref{tab:spectra}.

%				%%%%%%%%%%%%%%%%%%%%%%%%%%%
\begin{table}
\renewcommand{\tabcolsep}{0.20cm}
 \caption{Log of spectroscopic observations} 
  \label{tab:spectra}
%  \begin{center}
\begin{tabular}{ccccc}
\hline
   date    &  object &   BJD       &  exposure &  PA     \\
           &         &  246...     &     sec.  &$^\circ$ \\
\hline
\multicolumn{5}{c}{HRS}\\
2025-12-04 &  0      &  1014.23 &  3600     &   -     \\
2025-12-06 &  0      &  1016.15 &  3600     &   -     \\
2025-12-08 &  0      &  1018.28 &  3600     &   -     \\
2025-12-19 &  0      &  1029.25 &  7200     &   -     \\
2026-01-02 &  0      &  1043.21 &  7200     &   -     \\
\multicolumn{5}{c}{TDS} \\
2025-12-04 &  0      &  1014.26 &  600      &  70.8   \\
2025-12-06 &  0      &  1016.18 &  600      &  85.7   \\
2025-12-18 &  0      &  1028.19 & 1800      & $-32.0$   \\
2025-12-18 &  0      &  1028.17 & 1800      & 148.0   \\
2026-01-03 &  0      &  1044.19 &  600      &  67.5   \\
2025-12-20 &  1      &  1030.25 &  600      &  91.5   \\
2026-01-29 &  1      &  1070.19 & 2400      &  56.7   \\
2025-12-20 &  2      &  1030.25 &  600      &  91.5   \\
2025-12-18 &  HH     &  1028.23 &  900      & 158.0   \\
\hline
  \end{tabular} \\
%   \end{center}
The object designations follow those in Fig.~\ref{fig:Ha-image}. 
HH denotes Herbig-Haro objects.
\end{table}
% 				%%%%%%%%%%%%%%%%%%%%%%%%%%%

On December 3, 2025, we also obtained images of the {\IRAS} surroundings using the 4K$\times$4K CCD camera mounted on the 2.5-m telescope. Observations were carried out in the H$\alpha$ filter ($\lambda_{\text c}=656$~nm, $W=7.7$~nm, a total exposure time of $\Delta t=100$~min), and in the adjacent continuum filter H$\alpha$bc ($\lambda_{\text c}=643$~nm, $W=12$~nm, $\Delta t=50$~min).

            %%%%%%%%% End "Observations" section %%%%%%%%%%

                %%%%%%%%%%%%%%%%%%%%%%%%%%%%%%%%%%%%%%%%%%%%%%%%%%

\section{Results}
 \label{sect:results}

\subsection{Image of the surroundings of {\IRAS} in the H$\alpha$ line}
 \label{sec:Ha-image}

\begin{figure*}
%   \centering
 \includegraphics[width=\hsize]{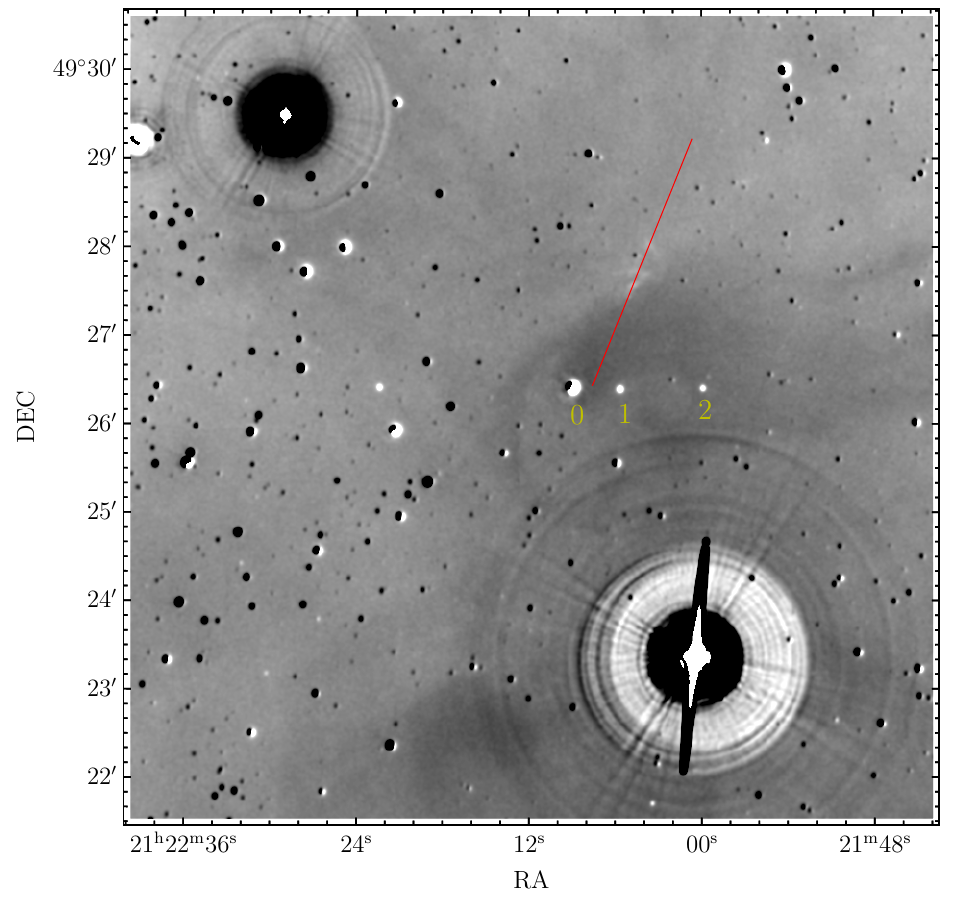}
\caption{Difference image of the {\IRAS} surroundings in the Halp and Halpbc filters. Objects 0, 1, and 2 mark {\IRAS}, star S1, and star S2, respectively. The red line segment passing through the Herbig-Haro objects (bright spots) indicates the position of the TDS spectrograph slit used to obtain the spectrum shown in Fig.~\ref{fig:jet-sp}. Further details are given in the text. 
 \label{fig:Ha-image}
}
\end{figure*}

Fig.~\ref{fig:Ha-image} shows the difference image of the 
{\IRAS} surroundings obtained in the Halp and Halpbc filters.
{\IRAS} is marked in the figure as object 0 at the southeastern
edge of the dark cloud D\,2944.
On the southern and southwestern edges of the same 
cloud, approximately 50{\arcsec} and 2\farcm27 west of {\IRAS},
lie two stars that appear as bright spots, indicating strong
H$\alpha$ emission in their spectra (see also 
\citealt{Barentsen-2014}). These stars -- 
Gaia DR3 2171042269468334080 and Gaia DR3 2171089101807867520 --
are labeled in Fig.~\ref{fig:Ha-image} as objects 1 and 2, 
respectively. Hereafter, we refer to them as stars S1 and S2.

%                           %%%%%%%%%%%%%%%%%%%%%%%%%%%
%                           %%%%%%%%%%%%%%%%%%%%%%%%%%%

\begin{figure}
   \includegraphics[width=\hsize]{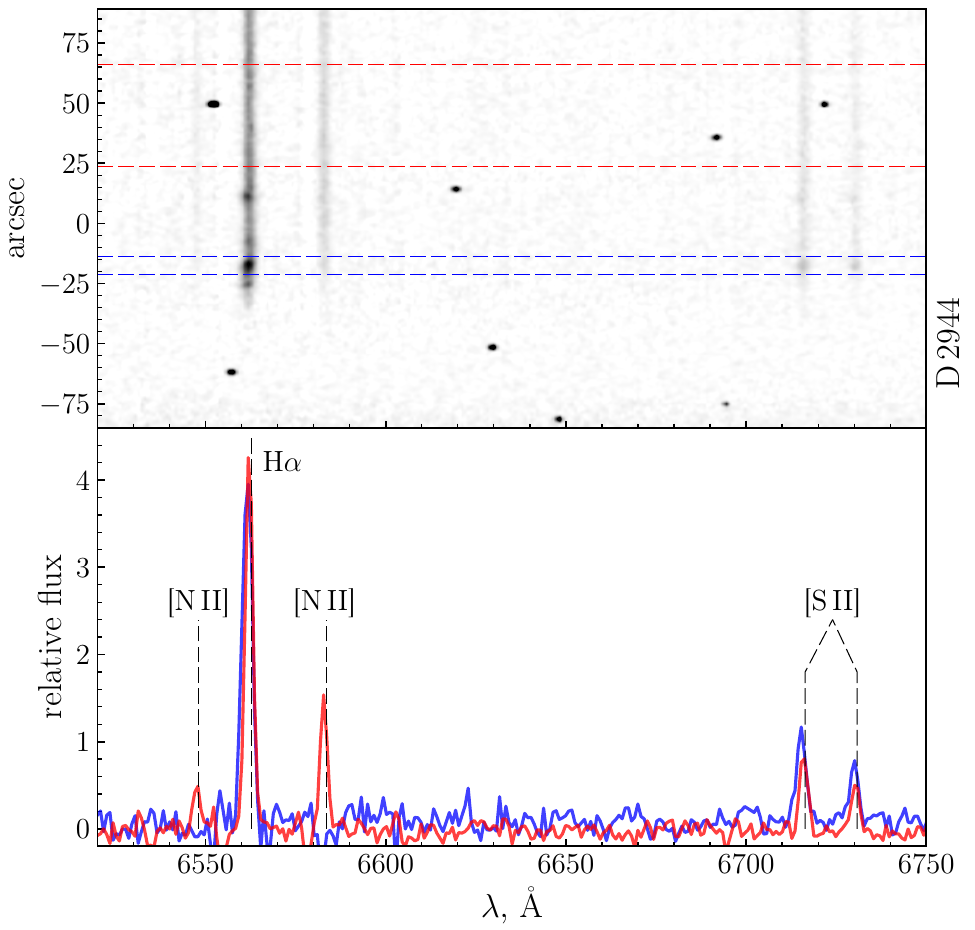}
\caption{2D (upper panel) and 1D (lower panel) TDS spectra of the emission nebulae located along the northern boundary of the dark cloud D\,2944. The red curve shows the spectrum of the background nebula within the region bounded by the red dashed lines, while the blue curve represents the spectrum of the brightest clump (within the blue dashed lines) with the background spectrum subtracted. Scattered black dots in the 2D image are cosmic-ray hits.  
  \label{fig:jet-sp}
}
\end{figure}

As can be seen from Fig.~\ref{fig:Ha-image}, a group of 
emission nebulae adjoins the northern part of the dark 
cloud D\,2944. These nebulae lie in the direction 
$\text{PA}_\text{HH} \approx -25\degr$ from {\IRAS} at the
projected distance of $d_\text{HH}\sim 2\farcm 5.$ 
According to the spectrum shown in Fig.~\ref{fig:jet-sp}, the 
brighter, irregularly shaped nebulae -- presumably Herbig-Haro 
(HH) objects -- are superimposed on an extended emission 
background. The spectra of the background gas and the HH objects 
differ in two key respects. First, the [\ion{N}{II}] doublet 
lines are present in the spectrum of the background gas but 
absent in that of the HH objects. Second, the barycentric 
radial velocity of the background gas is 
$V_{\text r}^{\text {bg}} \approx -40$~\kms, whereas that of
the brighter clumps is $V_{\text r}^{\text {HH}} \approx
-60$~\kms. In both the background and HH spectra,
the [\ion{S}{II}] $\lambda~6716$~\AA{} line is significantly 
stronger than the $\lambda~6731$~\AA{} line, implying an 
electron density $N_{\text e}\lesssim 200$~cm$^{-3}$ in the
line-emitting regions \citep{Proxauf-SII-ratio-2014}.

%               %%%%%%%%%%%%%%%%%%%%%%%%%%%
%               %%%%%%%%%%%%%%%%%%%%%%%%%%%

          %%%%%%%%%%%%%%%%%%%%%%%%%%%%%%%%%%%%%%%%

\subsection{Photometry}
 \label{sect:photometry}

\begin{figure}
   \centering
   \includegraphics[width=\hsize]{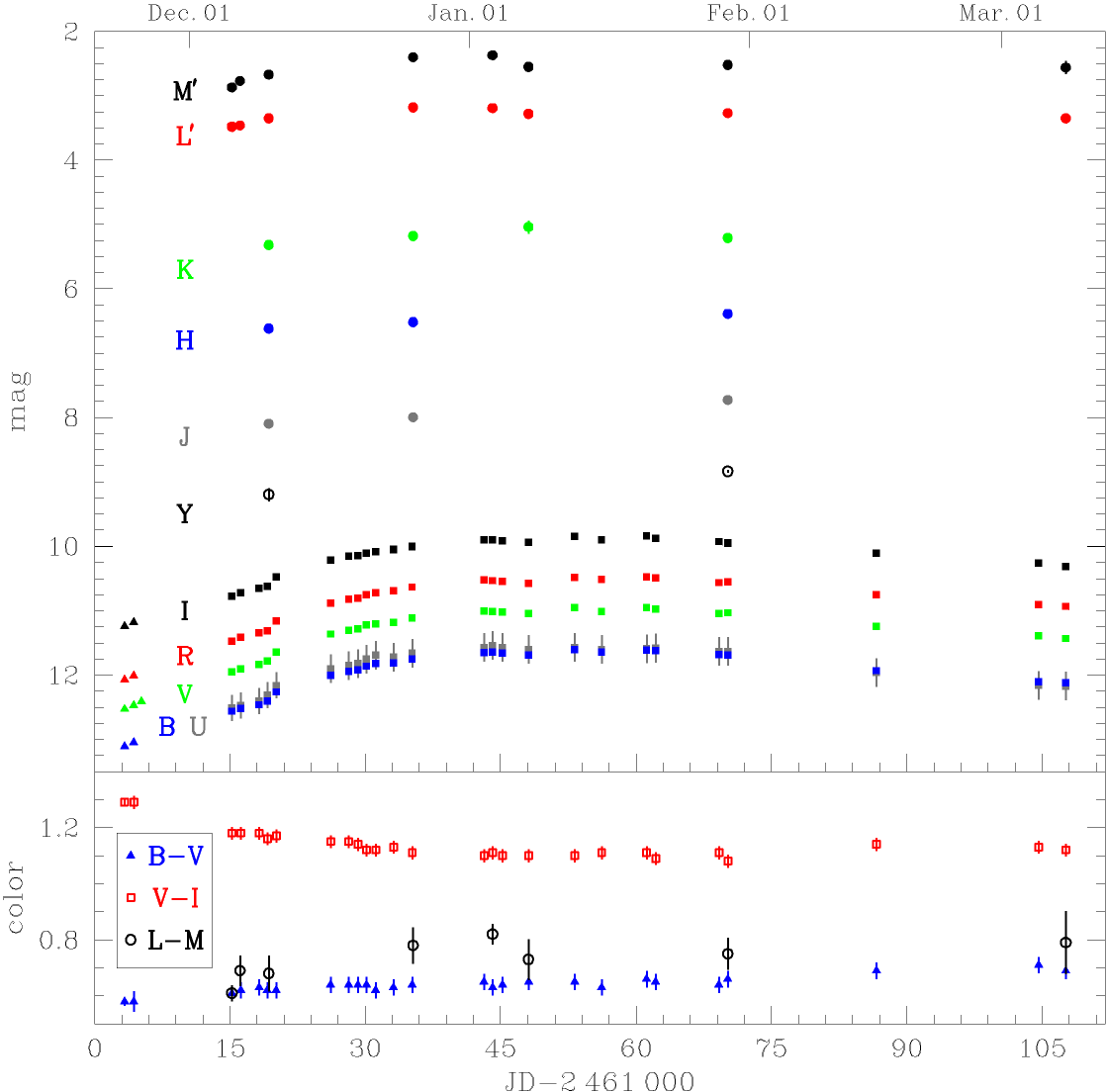}
\caption{Variation of the brightness and selected color indices of {\IRAS} in the optical and NIR bands. Data obtained prior to December 10, 2025 are taken from \citet{Kochkina-2025}.  
  \label{fig:LC-kgo}
}
   \end{figure}

Fig.~\ref{fig:LC-kgo} shows the evolution of the brightness of {\IRAS} in the optical and NIR bands during our observation period, along with several of its color indices. It is evident that the rapid brightening in the optical (by nearly $5^m$ during November 2025; \citealt{Stecklum-2025}) slowed down significantly in December. The stellar brightness reached maximum in January and then began to gradually decrease in optical band. Throughout our monitoring period, the source’s brightness variations from 0.37 to 4.8~\mcm{} occurred nearly simultaneously. More precisely, after December 10, 2025, the color indices $U-B \approx -0.08$ and $B-V \approx +0.63$ remained constant within the photometric uncertainties, while the $V-I$ color initially decreased slightly and then stabilized.

%                %%%%%%%%%%%%%%%%%%%%%%%%%%%
%                 %%%%%%%%%%%%%%%%%%%%%%%%%%%

\subsection{Polarimetry}
 \label{sect:polarim}

\begin{figure}
   \centering
   \includegraphics[width=\hsize]{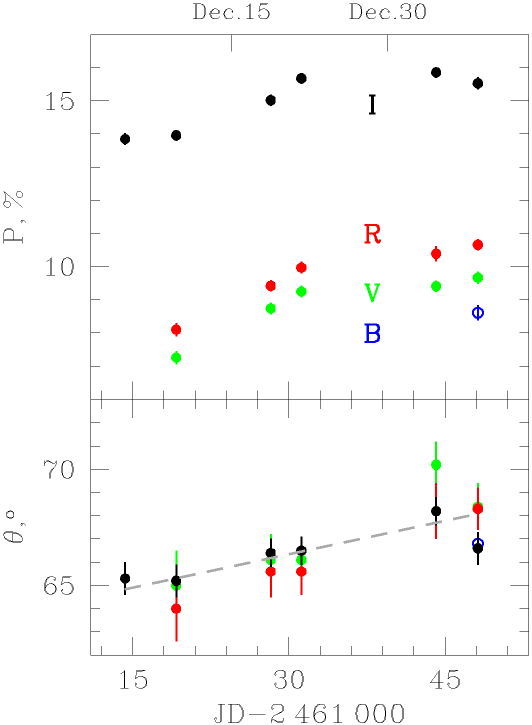}
\caption{Evolution of the degree of polarization $p$ and polarization angle $\theta$ of the radiation from {\IRAS}. The gray dashed line shows the least-squares fit. 
  \label{fig:PvsIWL}
}
   \end{figure}

It follows from Table~\ref{tab:polariz} and Fig.~\ref{fig:PvsIWL}
that the emission from {\IRAS} is strongly polarized in the $BVRI$
bands. At each epoch, the degree of polarization $p$ increases with
wavelength $\lambda$, while the polarization angle 
$\theta(\lambda)$ is constant within the measurement uncertainties.
During the course of our observations, the degree of polarization
in the $VRI$ bands increased, reaching $p\approx 16\%$ in the $I$
band. The polarization angle $\theta$ also appears to have 
increased slightly during the observation period. This 
conclusion is quantitatively supported by a weighted least-squares
fit (weights $1/\sigma^2_\theta$) to the time dependence 
$\theta=k\,t+\theta_0$, which yields a slope $k$ that differs
from zero at the level of $\approx 5\sigma_k$.

%               %%%%%%%%%%%%%%%%%%%%%%%%%%%
%               %%%%%%%%%%%%%%%%%%%%%%%%%%%

                %%%%%%%%%%%%%%%%%%%%%%%%%%%%

\subsection{Spectra}
 \label{sect:a-spectra}

\begin{figure*}
   \includegraphics[width=\hsize]{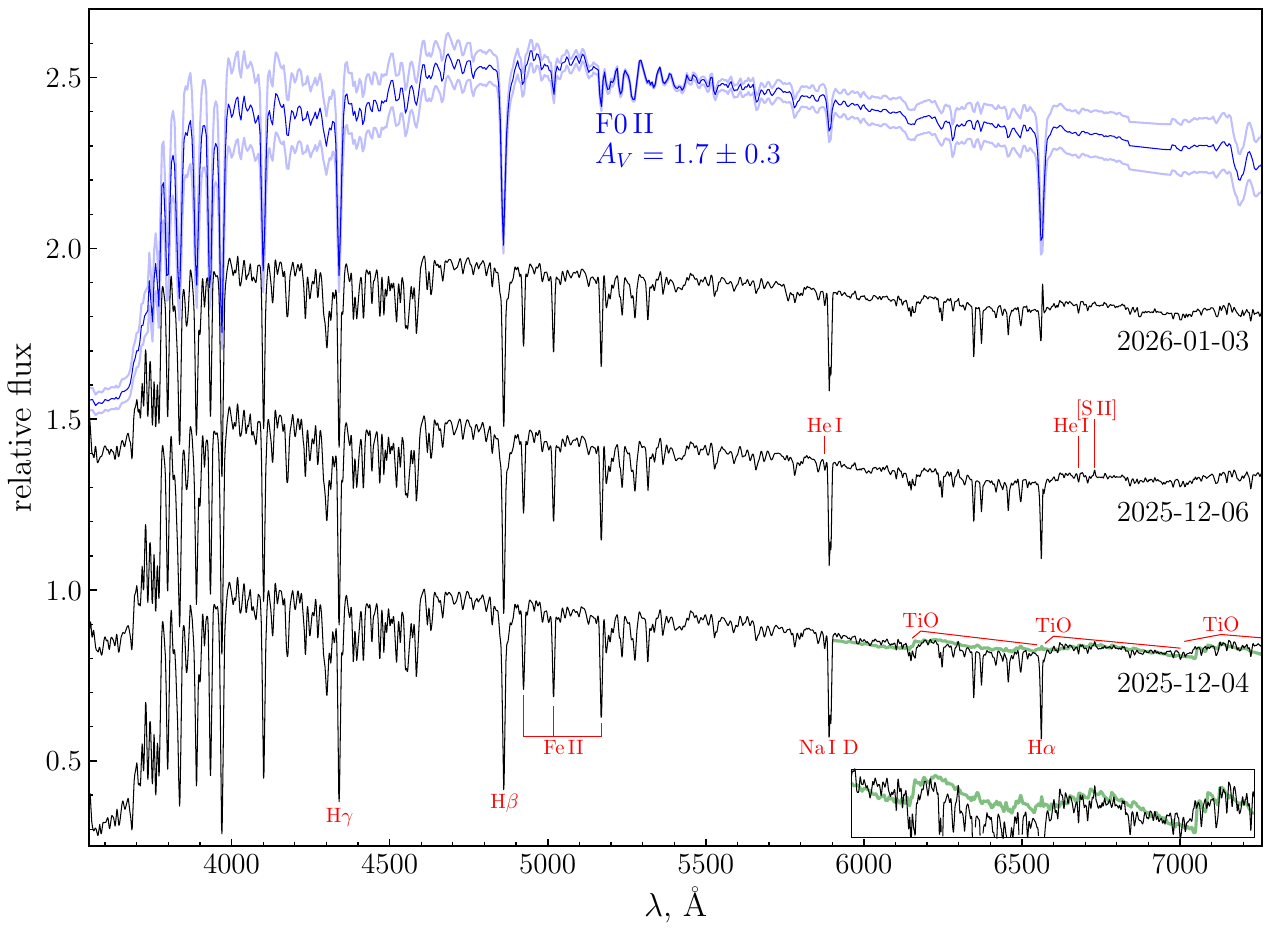}
\caption{Low-resolution spectra of {\IRAS} (black lines). The 
blue lines show the spectrum of an F0\,II star reddened by 
$A_{V}=1.7\pm 0.3.$ In the red part of the {\IRAS} spectrum, 
broad emission-like ``humps'' are present, which can be 
interpreted as TiO bands in emission. For comparison, the 
green line shows the inverted and scaled spectrum of the 
M-type star HD~211029; the same spectral regions are 
displayed at an enlarged scale in the insert. 
  \label{fig:TDS}
}
\end{figure*}

Fig.~\ref{fig:TDS} shows low-resolution spectra of {\IRAS} 
obtained with TDS on different dates. Overall, the spectrum
of the star resembles those of FUors \citep{Fischer-2023}: 
it is dominated by absorption features, including the 
\ion{Li}{I}~6708~\AA\ line, and exhibits both high-excitation
atomic lines (e.g. \ion{He}{I}, \ion{O}{I}, \ion{Si}{II}) 
and molecular TiO bands.

%                 %%%%%%%%%%%%%%%%%%%%%%%%%%%
%                   %%%%%%%%%%%%%%%%%%%%%%%

\begin{figure}
%   \centering
   \includegraphics[width=\hsize]{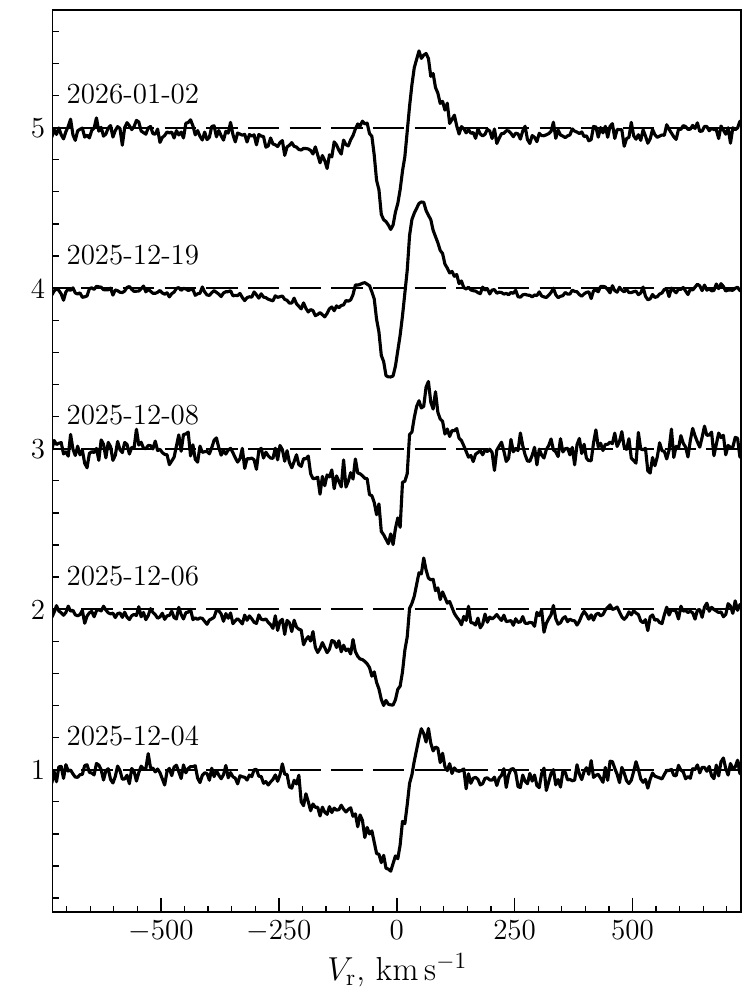}
\caption{Evolution of the H$\alpha$ line profiles. The 
profiles are vertically offset by steps of 1 for clarity.
The continuum level is indicated by a dashed line.
  \label{fig:Ha-prof}
}
\end{figure}

Another distinctive feature of the {\IRAS} spectrum that makes
it resemble those of FUors \citep{Herbig-1975, Fischer-2023} 
is the presence, in the blue wing of certain strong lines -- 
particularly H$\alpha$ (see Fig.~\ref{fig:Ha-prof}) -- of 
an extended absorption component. This fact indicates that 
the star, like FUors, has a gas shell 
that expands at velocities of $250-300$~{\kms}, i.e., a wind.

%                        %%%%%%%%%%%%%%%%%%%%%%%
%                          %%%%%%%%%%%%%%%%%%%%%%%

\begin{figure}
%   \centering
   \includegraphics[width=\hsize]{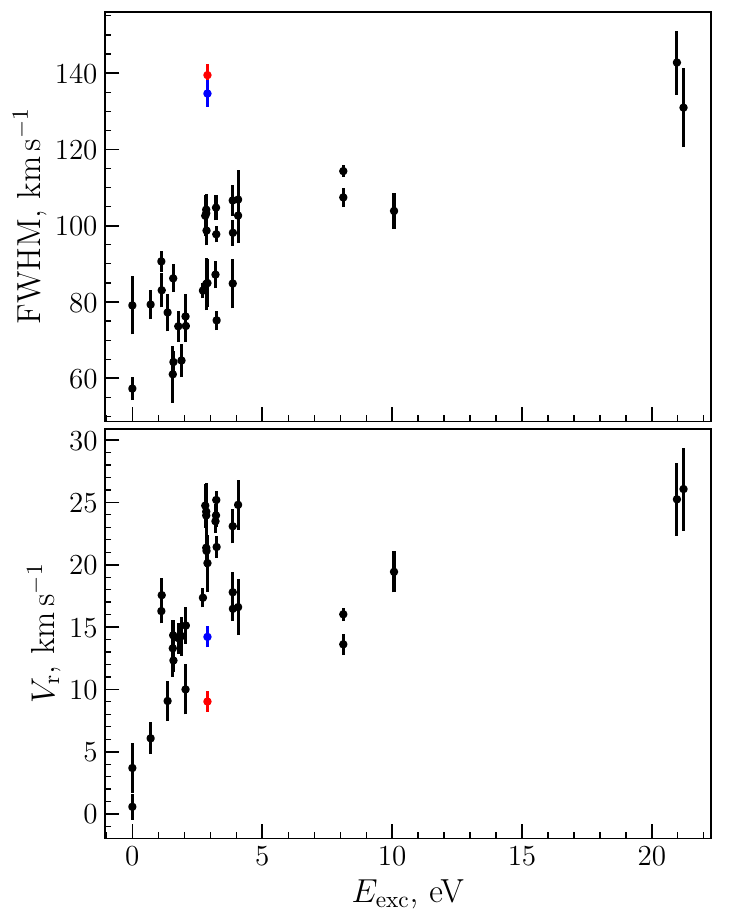}
\caption{Dependence of the line width and barycentric 
radial velocity $V_{\text r}$ on the excitation energy
of the lower level of the line. The two deviating points   
are strong \ion{Fe}{II} $\lambda\,4923.92$ (blue) and 
5018.44~\AA\ (red) lines.
  \label{fig:EVr}
}
\end{figure}

Nevertheless, the spectrum of {\IRAS} exhibits features 
that are atypical for FUors. First, as seen in 
Fig.~\ref{fig:TDS}, the TiO molecular bands appear in 
emission rather than absorption that has already been observed in the young eruptive stars LkH$\alpha$ 225S and V2429 Cyg as well as in IRAS 05451+0037 classified as Class I/II YSO \citep{Hillenbrand-2012, Hillenbrand-2022}. Second, there is a clear dependence both of the line width and radial velocity on the excitation energy (Fig.~\ref{fig:EVr}) -- a behavior not previously observed 
in classical FUors, to the best of our knowledge.

%                   %%%%%%%%%%%%%%%%%%%%%%%
%                   %%%%%%%%%%%%%%%%%%%%%%%

\begin{figure}
%   \centering
   \includegraphics[width=\hsize]{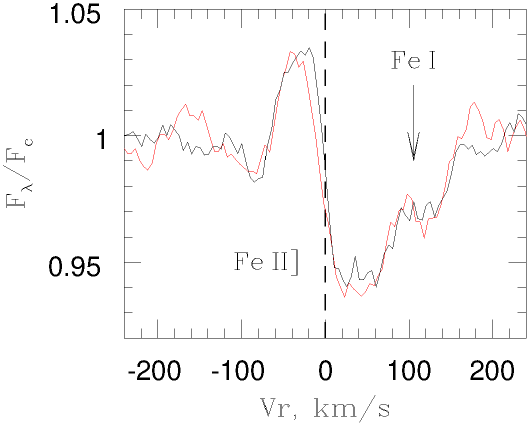}
\caption{Profiles of the \ion{Fe}{II} 6516.08\,\AA{} line in the spectra of {\IRAS} obtained on December 4 and 19, 2025 (red and black curves, respectively). The position of the \ion{Fe}{I}~6518.37~\AA{} line is indicated by an arrow. 
  \label{fig:FeII-line}
}
\end{figure}

The {\IRAS} spectrum exhibits several emission features. Firstly, 
as can be seen in Fig.~\ref{fig:EVr}, an emission component 
is present in the H$\alpha$ line profile in all the spectra 
obtained by us. Furthermore, two emission lines are visible 
in the stellar spectrum: [\ion{S}{II}]~6731~\AA{} (see 
Fig.~\ref{fig:TDS}) and [\ion{Ca}{II}]~7291~\AA{}. However, 
both are very weak, and the profile of the latter, as well 
as that of the [\ion{S}{II}]~6716~\AA{} line, is distorted 
by a neighboring absorption feature. Therefore, we refrain
from analyzing the detailed shapes of these lines, although 
we can state that the radial velocities of their centroids 
is $V_r<0.$ We also note that the 
[\ion{O}{I}]~5577, 6300, and 6363~\AA{} lines present 
in the {\IRAS} spectrum are of telluric origin. Another
emission feature, observed in all spectra of the star near 
$\lambda\approx 6516$~\AA{} (Fig.~\ref{fig:FeII-line}), 
probably corresponds to the fluorescent 
\ion{Fe}{II}]~6516.08~\AA{} line.\footnote{Transition 
$a\,^6S_{5/2} - z\,^6D^o_{7/2}$, whose upper level is pumped
by radiation from the $a~^6D_{9/2,\,7/2,\, 5/2}$ levels 
of the ground configuration of \ion{Fe}{II} 
\citep{Kramida-NIST-2024}.} To the best of our knowledge,
this line has not been previously reported in the spectra
of FUors, although it is present in the spectra of some
CTTS, e.g., RW~Aur \citep{Dodin-2012}.

%                    %%%%%%%%%%%%%%%%%%%%%%%
%                   %%%%%%%%%%%%%%%%%%%%%%%

                %%%%%%%%%%%%%%%%%%%%%%%%%%%%%%
                
\subsection{Radial velocity estimation}
 \label{sec:S12}
 
In the previous section we showed that the radial velocity 
of {\IRAS} varies with the excitation potential of the spectral
lines. To circumvent this complication and determine the stellar
radial velocity $V_{\rm r}$ , we proceeded as follows.

\begin{figure}
   \centering
   \includegraphics[width=\hsize]{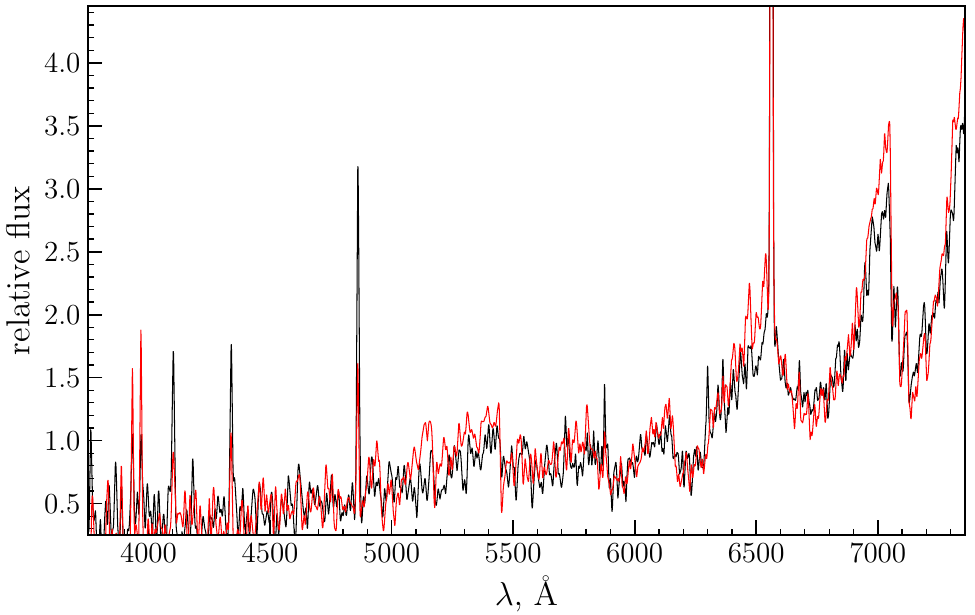}
\caption{TDS spectra of stars S1 (black line) and S2 
(red line). 
 \label{fig:S1S2}
}
\end{figure}

There are at least two CTTS in the vicinity of {\IRAS}, 
labeled as objects 1 and 2 in Fig.~\ref{fig:Ha-image}.
As shown in Table~\ref{tab:spectra}, we obtained two 
spectra of the star S1 and one spectrum of the star
S2. The spectra of both stars correspond to a late M 
spectral type, exhibit significant veiling, and 
display strong emission lines of hydrogen, 
\ion{Ca}{II}~H, K, and \ion{He}{I}.
[\ion{O}{I}] emission line is also present in the 
spectrum of the S1 star (Fig.~\ref{fig:S1S2}). For star S1, the 
equivalent width (EW) of H$\alpha$ line varied
from 100~\AA\ (December 20, 2025) to 55~\AA\ 
(January 29, 2026). In the spectrum of the S2 star, 
the H$\alpha$ line exhibits a double-peaked 
profile with ${\rm EW}\approx 53$~{\AA}.

             %%%%%%%%%%%%%%%%%%%%%%%%%%%%%%%%
%                
%
                %%%%%%%%%%%%%%%%%%%%%%%%%%%%%%%%

We have at our disposal a TDS spectrum of 
FN~Tau -- a CTTS of similar spectral type -- 
which we used as a template for determining 
$V_{\text r}$ of S1 and S2 stars. The radial 
velocity was measured in the wavelength range
$7040-7300$~\AA, where the most prominent 
absorption features -- TiO molecular bands -- 
are located. The observed spectra of S1 and S2 
were fitted to the FN~Tau template using a 
least-squares minimization with three free 
parameters: radial velocity, veiling, and 
a scaling factor to match the mean flux levels.
After correcting for the barycentric
radial velocity of FN~Tau, $V_{\text r}=+15.9$~{\kms} \citep{Jonsson-2020}, we obtained the following
barycentric radial velocities: for S1: $V_{\text r}=-7\pm 10$~{\kms} (December 20, 2025) and $V_{\text r}=+9\pm 5$~{\kms} (January 29, 2026); for S2: $V_{\text r}=+14\pm 6$~{\kms} (December 20, 2025). 

According to the Gaia DR3 catalogue \citep{Gaia-collaboration-2021}, both stars lie at the same 
distance from Earth ($d=500 \pm 5\,\%$~pc) and 
have similar proper motions $\mu_\alpha,~
\mu_\delta$ within the uncertainties.\footnote{
The catalogue does not provide distance or proper
motion data for {\IRAS}.} Based on these data, 
we conclude that S1, S2, and {\IRAS} are members
of a compact star-forming region associated with
the dark cloud D\,2944. In such a scenario, the
radial velocities of the stars in this region
should be similar. Therefore, our estimate
of $V_{\rm r}\approx +6\pm 10$~{\kms} for
{\IRAS} appears reasonable.

Note that the effective temperature of young M-type stars is $<3900$~K \citep{Herczeg-14}, implying that the masses of S1 and S2 stars are less than 0.5~M$_\odot$ \citep{Baraffe-2015}.
  
                %%%%%%%%%%%%%%%%%%%%%%%%%%%%%%
                
\subsection{Historical light curve}
\label{sec:LC-pg}{}

\begin{figure*}
\centering
\includegraphics[width=\hsize]{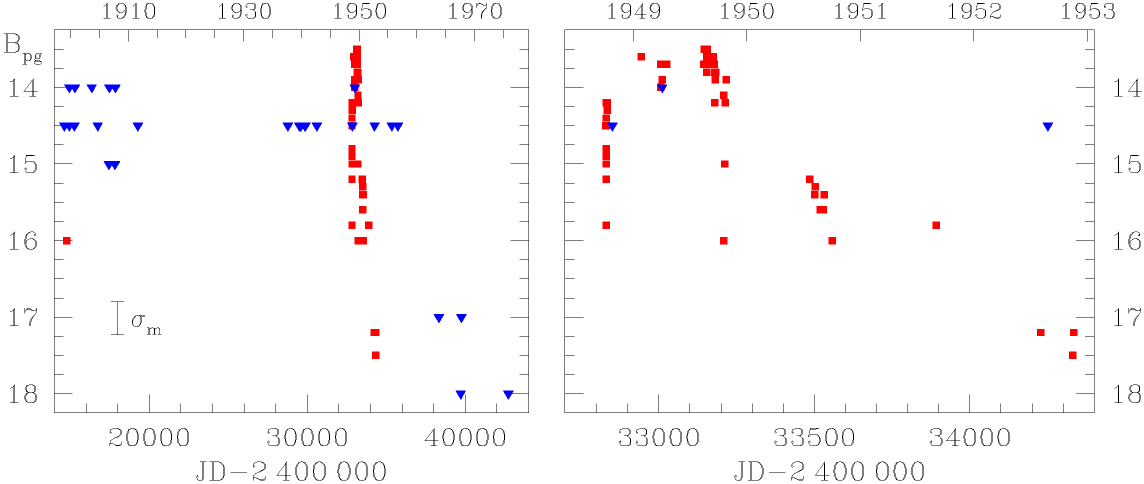}
\caption{Left panel: Light curve of {\IRAS} constructed from measurements of archival photographic plates obtained between 1899 and 1975. The measurement uncertainty is indicated in the lower-left corner (see text). Right panel: A zoomed-in segment of the light curve around the 1948 outburst. Red squares show measured magnitudes, while blue triangles denote upper limits. 
  \label{fig:LCpg}
}
\end{figure*}

To investigate the long-term variability of \IRAS, 
we estimated its brightness from archival photographic
plates in the SAI MSU collection (see Section~\ref{sect:observation}) and constructed a
century-long light curve, shown in Fig.~\ref{fig:LCpg}.

%            %%%%%%%%%%%%%%%%%%%%%%%%%%%
%           %%%%%%%%%%%%%%%%%%%%%%%%%%%

The star was first reliably detected on a 
photographic plate obtained on May 10, 1899 
at a magnitude of $m_{\rm pg}\approx 
16^{\rm m}.$ It remained undetected thereafter
until October 4, 1948, when it brightened to 
$m_{\rm pg}\approx 14\fm 5.$ Because {\IRAS} 
lies close to a dark nebula, suitable 
comparison stars were relatively distant, 
which reduced the photometric
accuracy. On the night of October 6-7, 1948, 
eleven photographic exposures of the field were 
obtained over a span of six hours, allowing
us to assess the precision of our visual 
magnitude estimates. The average uncertainty
is likely $\approx 0 \fm 3.$

As can be seen in Fig.~\ref{fig:LCpg}, the star’s brightness increased from October 1948 until mid-August 1949, after which it began to decline. The last detection of the star on the photographic plates occurred on November 17, 1952.
                
			%%%%%%%%% End of "Results section" %%%%%%%%%%

               %%%%%%%%%%%%%%%%%%%%%%%%%%%%%%%%%%

\section{Discussion} 
 \label{sect:discussion}
\subsection{{\IRAS} and HH objects}
 \label{sec:position}
The high degree of polarization of {\IRAS} and the 
wavelength independence of the polarization angle 
suggest that the polarization arises from scattering 
by dust. On p.~\pageref{sect:red-effect} we will present
evidence that this dust resides in an outflowing shell,
that is a dusty wind. If the wind is axially symmetric,
the polarization angle $\theta$ should be oriented 
perpendicular to the wind’s symmetry axis 
\citep{Dodin-2019}. Therefore, from Fig.~\ref{fig:PvsIWL}
it follows that the position angle of the wind axis 
is ${\rm PA}_{\rm w}=\theta \pm 90\degr = 156\degr 
\pm 1\degr$ or $-24\degr \pm 1\degr.$ 

The second value is in excellent agreement with the direction from {\IRAS} to the HH objects: $\rm{PA}_{HH}\approx -25\degr$ (see Fig.~\ref{fig:Ha-image}). We regard this alignment as strong evidence that the detected HH objects constitute part of a jet driven by the star under study. The barycentric radial velocity of the star is positive, whereas the HH objects exhibit negative radial velocities -- see Figs.~\ref{fig:EVr} and \ref{fig:jet-sp}, respectively. This difference implies that the HH objects are moving toward Earth while simultaneously receding from the star.

            %%%%%%%%%%%%%%%%%%%%%%%%%%%%%%%%%%%%%%%%%

\subsection{Extinction toward {\IRAS}}
 \label{sec:Av}

The absorption spectrum of {\IRAS} resembles those 
of A--F giants and supergiants, for 
example, V399~Car (A9\,I) and $\theta$~Sco (F1\,II).
This suggests that the spectral energy distribution 
(SED) in the optical band should also be consistent
with stars of these spectral types. We compared the observed SED of
{\IRAS} with the SED of A--F giants
and supergiants from the catalog of 
\citet{Pickles-1998} and found that, under the 
standard interstellar extinction law ($R_{\rm V}=
3.1$; \citealt{Cardelli-1989}), the extinction $\Av$ 
cannot exceed $2\fm 1$ (the value $2\fm 1$ is 
achieved for an A0\,I star). For the F0\,II 
spectrum -- the closest available match to A9\,I
and F1\,II in the catalog -- we obtain $\Av=1\fm7.$

For larger values of $\Av$, a kink appears near
4700~\AA{}, which is not observed in the spectrum
of {\IRAS} (Fig.~\ref{fig:TDS}). The kink becomes
even more pronounced when hotter stars
are used as spectral templates, ruling out the 
possibility of stronger reddening combined with
steeper dereddened continuum for {\IRAS}
than that of A--F giants. On the contrary, the 
presence of TiO emission bands indicates a 
contribution from radiation significantly cooler 
than that of A--F stars, which flattens the intrinsic
SED by enhancing its red portion. Accounting for 
this cool component would lead to a lower estimate
of $\Av.$

In the spectrum of {\IRAS} we found several diffuse interstellar bands 
(DIBs) from the list compiled by 
\citet{Herbig-1995}. It is reasonable to 
assume that these bands arise in the portion
of the dark nebula D\,2944 that partially
obscures the star. In particular, the EWs
of the $\lambda$~5797 and 6714~\AA{} DIBs
are measured to be $0.068
\pm 0.008$ and $0.097 \pm 0.012$~\AA, 
respectively, with barycentric radial 
velocities of $V_{\text r}= -11.1 \pm 1.7$
and $-15 \pm 2$~\kms, respectively.
Using the empirical relations between the 
EWs of these DIBs and extinction \citep{Lan-2015}, 
we derive $\Av$ in the range $0\fm 8-2\fm 1$,
accounting for the observed scatter in 
the correlations, and $A_V\approx 1\fm 3$ 
when using the mean calibration. We also note 
the absence of detectable interstellar
absorption in the \ion{Na}{I}~D and 
\ion{K}{I}~7699~\AA{} lines.

On the other hand, \citet{Stecklum-2025} reports 
that the {\it Swift} observatory did not detect 
X-ray emission from {\IRAS} in the $0.3-10$~keV 
band on November 27 and December 10, 2025, and 
on this basis concludes that $\Av=8^{\text m}.$
However, during our observation period the optical 
spectrum of the star was nearly flat 
(Figs.~\ref{fig:LC-kgo} and \ref{fig:TDS}). 
If such a high extinction were applied, the 
dereddened luminosity in the $0.4-5$~\mcm{}
range alone would reach $\sim 10^4$~L$_\odot$,
which is unrealistically large. This
strongly suggests that the $A_V=8^{\text m}$ 
estimate by \citet{Stecklum-2025} applies only 
to the region where the X-ray emission 
originates, which is obscured by the intense 
dusty wind as in the case of FU~Ori 
\citep{Skinner-2010}.

Thus, we adopt that the extinction in the 
direction of {\IRAS} $A_{\rm V}\lesssim 2^m,$ 
which is consistent with the estimate 
$\Av = 2\fm 17\pm 0\fm 22$ for D\,2944 nebula
as a whole \citep{Dobashi-2011}. For such 
extinction, the maximum expected 
$V$ band interstellar polarization would
be about 5.5\,\% \citep{Hiltner1956}, 
which is much less than the observed value.
Besides, interstellar polarization typically 
exhibits a maximum near $\lambda 
\approx 0.65$~\mcm. This gives reason to 
believe that the polarization of the object's
radiation occurs not in the interstellar 
medium, but in the dust shell of the star.

According to the catalog of 
\citet{Paegert-2021}, the $V$ band magnitude
of the star prior to the 2025 outburst was 
$19.28\pm 0.08.$\footnote{Thus, the amplitude of the outburst in the $V$ band exceeded $8^{\rm m}$.} 
Taking into account the adopted value of 
$A_{\rm V}$ and the distance to the star of 
500~pc, this corresponds to an absolute magnitude of $M_{\rm V}\approx 
8\fm 6.$ Since eruptive
YSOs are known to be younger than 10~Myr \citep{Fischer-2023}, evolutionary models
from \citet{Baraffe-2015} imply that only
stars with masses below 0.5~M$_\odot$ can
have absolute magnitudes fainter than 
this value of $M_{\rm V}.$

     %%%%%%%%%%%%%%%%%%%%%%%%%%%%%%%%%%%%%%

\subsection{Variability of the H$\alpha$ line profile} 
 \label{sect:Ha-prof}

We did not observe any significant differences
in the {\IRAS} spectra obtained on different
dates. The only exception is the profiles of 
the Balmer \ion{H}{I} lines, especially the 
H$\alpha$ line. We already noted (Sect.\,3.4)
that 
in all our spectra, starting from December 4,
the H$\alpha$ line 
has an emission component, which, however, 
is absent in the spectra obtained by 
\citet{Kochkina-2025} on November 25 and 27.
As can be seen from Fig.~\ref{fig:Ha-prof}, 
the H$\alpha$ line displays a P~Cyg profile,
in which the intensity of the redshifted 
peak $(V_{\rm r} >0)$ increases with time,
and the emission component in the blue
absorption wing near $V_{\rm r} \approx
-50$~\kms{} also strengthens. Note that 
\citet{Errico-2003} also 
observed the H$\alpha$ profile variations
in FU~Ori on timescales of several days,
and \citet{Powell-2012} even concluded that
these changes appear to be periodic.

              %%%%%%%%%%%%%%%%%%%%%%%%%%%%%

\subsection{The ``redshift'' effect of spectral lines}
\label{sect:red-effect}

It is reliably established that the spectrum 
of a FUor is formed in the protoplanetary disk, 
whose accretion luminosity greatly exceeds the
luminosity of the central star \citep{Fischer-2023}. Within this framework, the 
presence of lines with very different excitation potentials
$E_{\rm exc}$ in the FUor spectra
is explained by the 
radial decrease of the effective temperature of 
the disk with increasing distance from the rotation
axis $r$. Since the rotational velocity also 
decreases with $r$, the model predicts that the 
observed line width -- e.g. the full width at half 
maximum (FWHM) -- should be smaller for lines with 
lower $E_{\rm exc}$ \citep{Hartmann-Keyon-1996}.

Meanwhile, \citet{Herbig-1989} already noted that,
for all FUors known at the time, no dependence of
${\rm FWHM}$ on excitation energy $E_{\rm exc}$ 
was observed, at least in the visible
range -- see also \citet{Herbig-2003, Petrov-2008}.
The lack of correlation between ${\rm FWHM}$ and 
$E_{\rm exc}$ may indicate that the regions of the
disk responsible for optical lines formation 
do not follow Keplerian rotation and/or that
the FUor wind significantly distorts the observed
line profiles \citep{Petrov-2008}.

We will not discuss these reasons in further 
detail, because in the case
of {\IRAS} a clear correlation between ${\rm FWHM}$ and
$E_{\rm exc}$ is indeed present. Moreover, 
as shown in Fig.~\ref{fig:EVr}, 
the radial velocity $V_{\rm r}$ of the lines also
increases with $E_{\rm exc}$, a behavior not 
predicted by standard disk accretion models.

It can be seen from Fig.~\ref{fig:EVr} that both
dependencies arise because
the red wing
of the line becomes increasingly extended with higher
$E_{\rm exc}$. We believe that the change in line 
profiles is associated with the scattering of 
radiation \citep{Grinin-2012}, a process whose significant
role in shaping the spectrum of \IRAS{} is strongly
supported by the high degree of polarization observed.

%                   %%%%%%%%%%%%%%%%%%%%%%%
\begin{figure}
%   \centering
\includegraphics[width=\hsize]{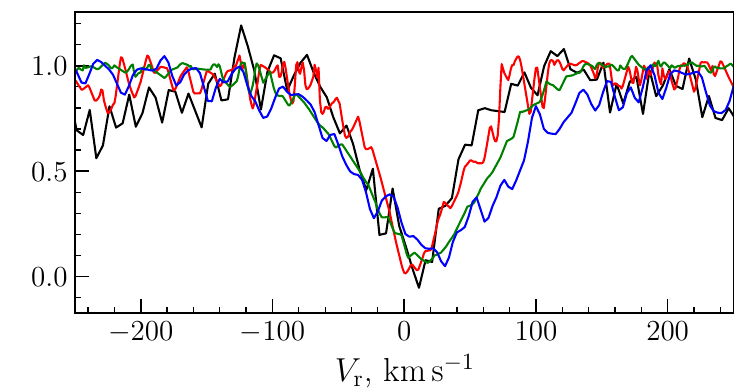}
\caption{Dependence of the line profile shape on the 
excitation energy $E_{\rm exc}$ of the lower level.
The black curve shows the \ion{Ba}{II} 6141.71~\AA{} 
line ($E_{\rm exc}=0.7$~eV); the red curve represents
the average profile of the \ion{Sc}{II} 5031.02, 5526.79,
and 5641.00~\AA{} lines (1.4--1.8~eV); the green curve
is the average profile of the \ion{Si}{II} 6347.11 and
6371.37~\AA{} lines (8.1~eV); and the blue curve shows
the average profile of the \ion{He}{I} 5875.66 and 
6678.16~\AA{} lines lines (21~eV). Barycentric 
velocity is plotted along the abscissa. 
  \label{fig:lp}
}
\end{figure}
%                   %%%%%%%%%%%%%%%%%%%%%%%

To illustrate this effect, consider a dusty wind in 
which the material moves radially outward at a 
constant velocity $V_{\rm w}$. A photon emitted by
the central star at frequency $\nu_0$ will be 
Doppler-shifted in the rest frame of a dust grain
to $\nu_1=\nu_0 (1-V_{\rm w}/c),$ where $c$ is 
the speed of light. After a single coherent 
scattering event, the photon reaches the observer
with frequency 
$\nu_2=\nu_1(1+V_{\rm w} \cos \beta/c),$ 
where $\beta$ is the angle between the radius 
vector pointing from the dust grain to the star 
and the line of sight. Eventually, the original wavelength 
$\lambda_0$ is shifted by an amount
\begin{equation}
\frac{\Delta \lambda}{\lambda_0} = - \frac{\Delta \nu}{\nu_0}
\approx \left(1-\cos\beta \right)\, \frac{V}{c}  \geqslant 0,
 \label{eq:redsift}
\end{equation}
toward longer wavelengths. The effects of line
shift and broadening due to {\it multiple} 
scattering are discussed in detail by 
\citet{Grinin-2006}.

The dependence of the line width on the 
excitation potential can be associated not 
only with the thermal structure of the 
accretion disk, but also with differences 
in the scattering parameters for disk 
regions with different temperatures.
Different scattering geometries and/or 
scattering optical depth lead to different
distortions of spectral lines. In other words,
the observed spectrum depends on the geometry
and kinematics of both the emitting region 
and the dusty wind, as well as on dust 
properties \citep{Grinin-2012}. 

Having no information on these 
parameters, as well as spectropolarimetric data
for {\IRAS}, we are forced to limit ourselves 
to a qualitative interpretation of the 
``redshift'' effect.

             %%%%%%%%%%%%%%%%%%%%%%%%%%%%%%%%%%%%%%
 
\subsection{Accretion rate estimate} 
 \label{sec:Mdot}
The phenomenological division of eruptive YSOs into
FUors and EXors, described in the Introduction, 
reflects fundamental differences in the accretion 
regimes of the protoplanetary disk's matter onto the
central star \citep{Hartmann-2016}. EXors exhibit 
so-called magnetospheric accretion, similar to that in 
CTTSs, but with a much higher accretion rate $\dot 
M_{\rm ac}.$ For accretion rates typical of FUors 
($\dot M_{\rm ac}\gtrsim 10^{-5}$~M$_\odot$yr$^{-1}$),
the protoplanetary disk “pushes through” the stellar
magnetic field and makes direct contact with
the stellar surface. Consequently, energy is 
released in a boundary layer near the star 
rather than in an accretion shock, as is the case for 
EXors and CTTSs.

\begin{figure}
   \includegraphics[width=\hsize]{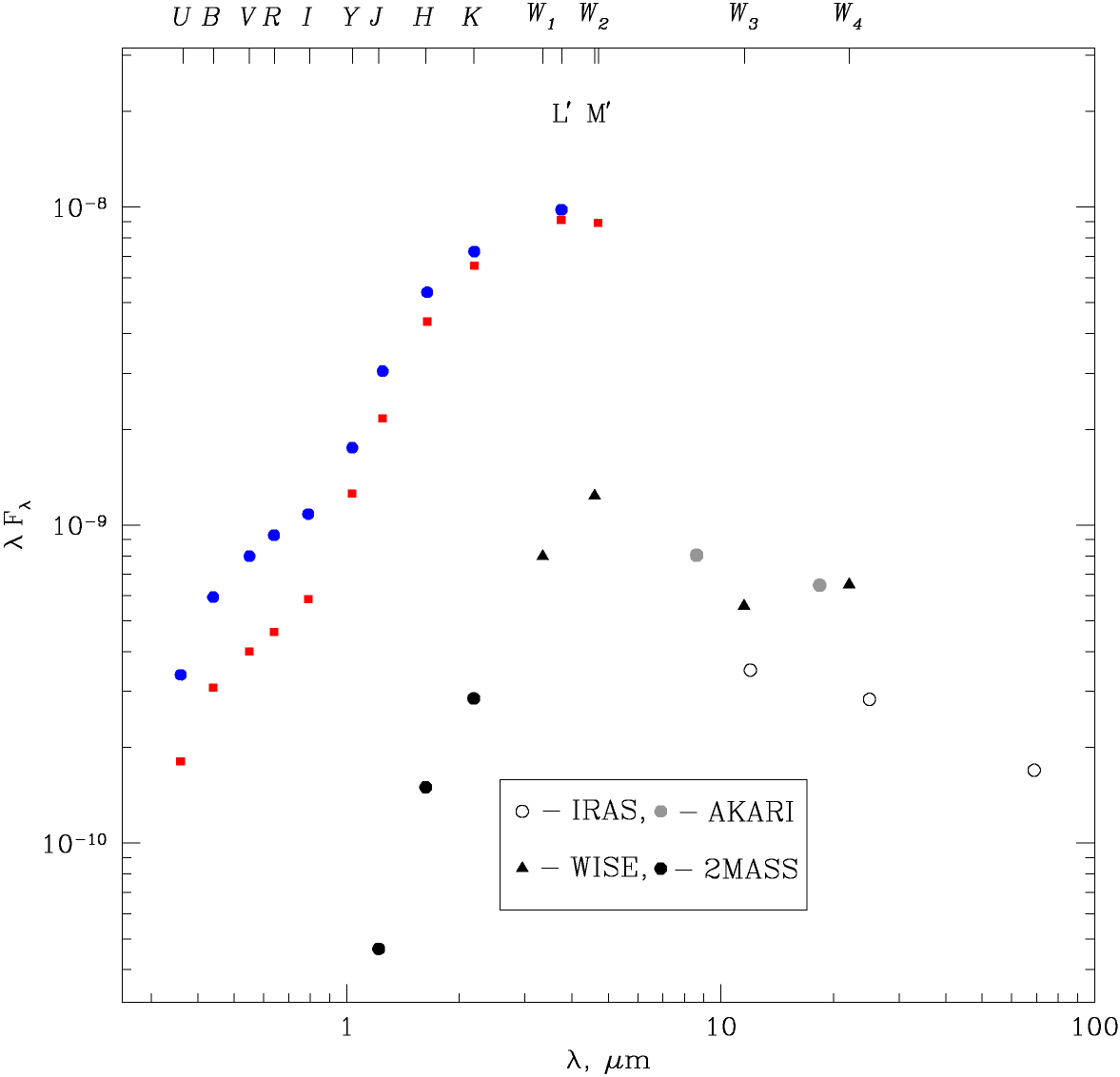}
\caption{Temporal evolution of the SED of {\IRAS} without
extinction correction. Red and blue symbols show the SED 
on December 9, 2025 and January 29, 2026, respectively. 
Pre-outburst data were obtained at different epochs and 
compiled from various sources; see text for details. 
  \label{fig:sed-iras}
}
\end{figure}

Fig.~\ref{fig:sed-iras} shows the SED of {\IRAS} in
the $0.4-5$~\mcm{} range, uncorrected for extinction.
The SED is constructed from photometric data obtained
at the beginning (December 9, 2025) and the middle (January
29, 2026) of our observations (see Tables~\ref{tab:Optph}
and \ref{tab:IRph}). During this period, the total flux
in the considered wavelength range increased by 
approximately $15$\,\%, reaching a value of 
$\approx 1.1\times 10^{-8}$~erg s$^{-1}$cm$^{-2}$.
At a distance of 500\,pc, this corresponds to
a luminosity of $L\approx 90$~L$_\odot.$

%                           %%%%%%%%%%%%%%%%%%%%%%%%%%%
%                           %%%%%%%%%%%%%%%%%%%%%%%%%%%

For comparison, Fig.~\ref{fig:sed-iras} also 
shows the pre-outburst SED of the star, 
constructed from archival data obtained at 
different epochs from the 2MASS 
\citep{Skrutskie-2006}, WISE \citep{Furlan-2011},
AKARI \citep{Ishihara-2010}, and IRAS 
\citep{Neugebauer-1984} surveys. From these 
data we find that the pre-outburst bolometric
luminosity of the star was $\approx 20$~L$_\odot,$
with the range $0.4-5$\,\mcm{} accounting for only 
$\sim 10$~\%{}. It is unlikely that the flux 
at wavelengths $\lambda >5$\,\mcm{} remained 
unchanged during the outburst. Therefore, it 
is reasonable to assume that the bolometric 
luminosity during this period was at least 
twice the value inferred from the $\lambda <
5$\,\mcm{} flux alone, reaching $L_* \sim 
200$~L$_\odot.$

If this luminosity is powered by accretion 
onto a star of mass $M_*$ and radius $R_*,$
then from the relation $L_*=\dot M_{ac}GM_*/R_*$ \citep{Shakura-1973} we obtain that the 
accretion rate at that time must have exceeded 
\begin{equation}
\dot M_{\rm ac} = 6\times 10^{-6}
\, \left( \frac{R_*}{R_\odot} \right)
\left( \frac{M_\odot}{M_*} \right)\, 
{\rm M}_\odot\,\text{yr}^{-1}.
\end{equation}
According to \citet{Baraffe-2015}, for young stars
younger than 10~Myr with masses $M_*<0.5$~M$_\odot$
(see p.~\pageref{sec:Av}), the ratio 
$(R_*/R_\odot)/(M_*/M_\odot) > 5.$ Therefore, we 
conclude that during our observation period the 
accretion rate was $\gtrsim 3\times 
10^{-5}$~M$_\odot$\,yr$^{-1}$.

             %%%%%%%%%%%%%%%%%%%%%%%%%%%%%%%%%%%%%%

\subsection{Activity of the star before the 2025 outburst }
 \label{sec:fuor-like}

\citet{Kochkina-2025} found no evidence in the 
ASAS-SN data  for strong outbursts of {\IRAS} between 
2015 and October 2025. However, \citet{Stecklum-2025}, 
relying on data from the Gaia \citep{Mowlavi-2021} 
and NEOWISE \citep{Mei-2023} observations, reports 
that before the outburst, {\IRAS} experienced 
long-term brightness variations of small amplitude
in both the visible (after 2015) and NIR 
(after 2010) spectral regions. \citet{Stecklum-2025}
also noted that a strong H$\alpha$ line was 
observed in the spectrum of the star before 2025
\citep{Barentsen-2014} and suggested that intense 
accretion of circumstellar matter was already occurring before the 2025 outburst. This is further supported by the shape of the SED 
(Fig.\,\ref{fig:sed-iras}) and the high 
luminosity of the star during that period.

Finally, we note (Section~3.6)
that {\IRAS} underwent a prolonged brightness
increase in the past. As can be seen 
from Fig.\,\ref{fig:LCpg}, the star 
experienced an outburst in 1948, during which
its brightness rose to $m_{\rm pg}\approx
13\fm 5,$ and then declined over approximately 
three years to a level of $m_{\rm pg}
\approx 18^{\rm m}.$ If this level 
represents the pre-outburst state of 
the star, then the amplitude of the 1948 outburst
was $\Delta m_{\rm pg}\approx 4\fm 5.$

It's difficult to say how similar the 1948 and 2025 
outbursts are: we do not know what the star's spectrum 
was during the first outburst, how quickly its 
brightness increased in 1948, and how long it 
will continue to decline after the second outburst.
Already now, {\IRAS} is approximately 
$2^{\rm m}$ brighter in the $B$ band than it was in 1949, however, we cannot 
rule out that the true maximum of the 1948 outburst was missed --- see 
Fig.\,\ref{fig:LCpg}. Nevertheless, the very fact that
an outburst of comparable amplitude occurred in 
an object that currently exhibits many characteristic features of a FUor is highly significant.

\citet{Herbig-1977}, based on statistical 
grounds, concluded that the FUor phenomenon
in young stars should be recurrent -- see also
\citet{Hillenbrand-2015}. Meanwhile, among 
eruptive YSOs, considered ``bonafide'' FUors in the \citet{Contreras-2025}
catalog, no confirmed cases of repeated outbursts have been observed to date. On
the other hand, in a number of objects with 
{\it some} characteristic FUor-like features -- 
for example, V1647~Ori \citep{Aspin-2009} or 
V899~Mon \citep{Ninan-2015} -- the brightness 
dropped sharply a few years after the maximum,
and then began to increase again. Unlike these 
objects, {\IRAS} had a much longer interval 
between outbursts -- at least 77 years.

Not all mechanisms invoked to explain FUor outbursts 
(see Introduction) can account for this kind of phenomenon.
However, some models explicitly predict the 
presence of ``double'' outbursts in FUors, e.g., 
\citet{Grigoryev-2025}. In this regard, we recall 
that in EXors, repeated prolonged outbursts are 
observed quite frequently -- see, for example, 
\citet{Herbig-1977, Dodin-2016, Giannini-2015}.

  %%%%%%%%%%  End of "Discussion" section %%%%%%%%%%

\section{Conclusion}
 \label{sect:summary}

There is no doubt that {\IRAS} is an eruptive 
YSO in the transitional stage from a protostar 
to a young star. The object is embedded within
the dark nebula D\,2944, which is a compact 
star-forming region located at a distance of 
approximately 500~pc and hosting low-mass YSOs
$(M\lesssim 0.5$~M$_\odot)$. We estimate the
extinction toward {\IRAS} to be $A_{\rm V} 
\lesssim 2^{\rm m}.$

The outburst of {\IRAS}, which began in 
late October 2025, resembles FUor-type 
eruptions -- phenomena driven by a dramatic
increase in the accretion rate of material
from the protoplanetary disk onto the 
forming star -- in several aspects:
\begin{itemize}
 \item A large outburst amplitude $(\Delta
 V> 6^{\rm m})$ corresponding to an accretion
 rate of $\gtrsim 3\times 10^{-5}$~M$
 _\odot$yr$^{-1}$;
 \item An absorption spectrum in the range
$0.4 - 0.75$~\mcm{} resembling those of A--F 
giants and supergiants, combined with the
presence of TiO molecular bands;
 \item An outflowing shell expanding at 
 velocities up to $\approx 300$~\kms{}, 
 in which the H$\alpha$ line with a P~Cyg
 profile is formed;
 \item A group of HH objects in the 
 vicinity of {\IRAS}, receding from it 
 with radial velocities of $V_{\rm r}
 \approx 70$~\kms.
\end{itemize}

At the same time, we have identified several 
distinctive features in {\IRAS}. As the star 
brightened, its radiation became increasingly
polarized, reaching $\approx 16$~\%{} in the
$I$ band by the end of our observations. We 
attribute this to the presence of a dusty 
wind. The scattering of the central source's
radiation by dust grains produces a 
dependence of both the FWHM and $V_{\rm r}$ 
of spectral lines on their excitation 
potential -- an effect not previously 
observed in FUors. Another notable feature
of {\IRAS} is an outburst of comparable 
amplitude that occurred 77 years ago.

We also discovered for the first time an 
emission line at $\lambda = 6516$~\AA{} and
TiO molecular emission bands in the 
spectrum of an eruptive YSO with an 
absorption-dominated continuum. It is 
difficult to assess how critical these emission 
features are for understanding the outburst 
mechanism of {\IRAS} and other young eruptive stars. However, we note that the
CO bands near $\lambda = 2.3$~\mcm{} -- 
typically seen in absorption in classical 
FUors -- are observed in emission in the 
eruptive YSOs V1647~Ori and IRAS~06297+1021W \citep{Connelley-2018}, which led 
\citet{Contreras-2025} to classify them as 
intermediate objects between FUors and EXors.

It is still premature to classify {\IRAS} 
as a ``bonafide'' FUor, since -- by definition 
-- this group includes only objects whose 
brightness fades significantly over timescales
of years to decades after maximum light
\citep{Samus-2017}. However, the issue is
not merely one of formal classification of
eruptive YSOs, but rather of understanding
the physical mechanisms responsible for
the diversity of their observed properties.
The nontrivial features exhibited by {\IRAS}
provide strong grounds to assert that this
young star warrants close attention. It would
be highly desirable not only to repeat the
observations presented here, but also to 
obtain spectropolarimetric data and to monitor
the evolution of its SED at wavelengths beyond
5~{\mcm}.

 %%%%%%%%  End of "Conclusion" section %%%%%

\section*{Acknowledgements}

We thank the staff of the CMO SAI MSU for their assistance during the observations. We are also grateful to M.~R.~Gilfanov, D.~A.~Lashin, and A.~N.~Tarasenkov for valuable discussions. We gratefully acknowledge the use of data from the following databases in this work: SIMBAD (CDS, Strasbourg, France), the Astrophysics Data System (NASA, USA), the NIST Atomic Spectra Database (NIST, USA), and the Atomic Line List (University of Kentucky, USA).

The study was conducted under the state assignment of Lomonosov Moscow State University.

			%%%%%%%%%%%%%%%%%%%%%%%%%%%%%%
            
\section*{Data availability}
The results of the photographic magnitude measurements are available upon request.

			%%%%%%%%%%%%%%%%%%%%%%%%%%%%%%           

\bibliographystyle{aspb1}
\bibliography{iras}

\section*{Appendix}
\label{sect:addition}
\renewcommand{\thetable}{A1}
\begin{table*}
\renewcommand{\tabcolsep}{0.2cm}
 \caption{Optical photometry for {\IRAS}}
  \label{tab:Optph}
  \begin{tabular}{ll|ll|ll|ll|ll|ll}
\hline
Date & UT &$U$ &$\sigma_U$ &$B$ &$\sigma_B$ &$V$ &$\sigma_V$ &$R$ &$\sigma_R$ 
& $I$ & $\sigma_I$ \\
\hline
05.12.2025 & 17:21 &12.51 & 0.22 &12.56 &0.02  &11.95 &0.02  &11.47 &0.02  &10.77 &0.01  \\ 
06.12.2025 & 16:46 &12.47 & 0.22 &12.52 &0.02  &11.90 &0.02  &11.41 &0.02  &10.72 &0.01  \\
08.12.2025 & 17:45 &12.40 & 0.22 &12.46 &0.02  &11.83 &0.02  &11.34 &0.02  &10.65 &0.01  \\
09.12.2025 & 15:27 &12.31 & 0.22 &12.40 &0.02  &11.78 &0.02  &11.31 &0.02  &10.62 &0.01  \\
10.12.2025 & 15:14 &12.16 & 0.22 &12.26 &0.02  &11.64 &0.02  &11.16 &0.02  &10.47 &0.01  \\
16.12.2025 & 16:00 &11.90 & 0.22 &12.00 &0.02  &11.36 &0.02  &10.88 &0.02  &10.21 &0.01  \\
18.12.2025 & 16:03 &11.85 & 0.22 &11.94 &0.02  &11.30 &0.02  &10.82 &0.02  &10.15 &0.01  \\
19.12.2025 & 16:34 &11.82 & 0.22 &11.92 &0.02  &11.28 &0.02  &10.80 &0.02  &10.14 &0.01  \\
20.12.2025 &16:15  &11.75 &0.22  &11.86 &0.02  &11.22 &0.02  &10.75 &0.02  &10.10 &0.01  \\
21.12.2025 &15:42  &11.69 &0.22  &11.82 &0.02  &11.20 &0.02  &10.72 &0.02  &10.08 &0.01  \\
23.12.2025 &15:04  &11.72 &0.22  &11.81 &0.02  &11.18 &0.02  &10.69 &0.02  &10.05 &0.01  \\
25.12.2025 &16:40  &11.66 &0.22  &11.75 &0.02  &11.11 &0.02  &10.63 &0.02  &10.00 &0.01  \\
02.01.2026 &15:59  &11.57 &0.22  &11.65 &0.02  &11.00 &0.02  &10.52 &0.02  & 9.90 &0.01  \\
03.01.2026 &14:53  &11.54 &0.22  &11.64 &0.02  &11.01 &0.02  &10.53 &0.02  & 9.90 &0.01  \\
04.01.2026 &17:01  &11.57 &0.22  &11.66 &0.02  &11.02 &0.02  &10.54 &0.02  & 9.92 &0.01  \\
07.01.2026 &14:55  &11.60 &0.22  &11.69 &0.02  &11.04 &0.02  &10.57 &0.02  & 9.94 &0.01  \\
12.01.2026 &16:10  &11.57 &0.22  &11.60 &0.02  &10.95 &0.02  &10.48 &0.02  & 9.85 &0.01  \\
15.01.2026 &16:49  &11.60 &0.22  &11.64 &0.02  &11.01 &0.02  &10.51 &0.02  & 9.90 &0.01  \\
20.01.2026 &15:10  &11.59 &0.22  &11.61 &0.02  &10.95 &0.02  &10.47 &0.02  & 9.84 &0.01  \\
21.01.2026 &15.16  &11.58 &0.22  &11.62 &0.02  &10.97 &0.02  &10.49 &0.02  & 9.88 &0.01  \\
28.01.2026 &15:32  &11.63 &0.22  &11.68 &0.02  &11.04 &0.02  &10.56 &0.02  & 9.93 &0.01  \\
29.01.2026 &15:24  &11.63 &0.22  &11.69 &0.02  &11.03 &0.02  &10.55 &0.02  & 9.95 &0.01  \\
15.02.2026 &03:09  &11.96 &0.22  &11.93 &0.02  &11.24 &0.02  &10.75 &0.02  &10.10 &0.01 \\
05.03.2026 &02:27  &12.16 &0.22  &12.10 &0.02  &11.39 &0.02  &10.90 &0.02  &10.26 &0.01 \\
08.03.2026 &02.44  &12.17 &0.22  &12.12 &0.02  &11.43 &0.02  &10.93 &0.02  &10.31 &0.01 \\
\hline
 \end{tabular} \\
\end{table*}
% 		         %%%%%%%%%%%%%%%%%%%%%%%%%%%
%				%%%%%%%%%%%%%%%%%%%%%%%%%%%
\renewcommand{\thetable}{A2}
\begin{table*}
\renewcommand{\tabcolsep}{0.2cm}
 \caption{IR photometry for \IRAS}
  \label{tab:IRph}
  \begin{tabular}{ll|ll|ll|ll|ll|ll|ll}
\hline
Date & UT &$Y$ &$\sigma_Y$ &$J$ &$\sigma_J$ &$H$ &$\sigma_H$ &$K$ &$\sigma_K$ 
& $L^{\prime}$ & $\sigma_L$ & $M^{\prime}$ & $\sigma_M$ \\
\hline
05.12.2025 & 17:10 &     &      &     &     &     &     &     &     &3.48 &0.02 &2.87 &0.02 \\
06.12.2025 & 14:40 &     &      &     &     &     &     &     &     &3.46 &0.02 &2.77 &0.05 \\
09.12.2025 & 19:05 &9.2  & 0.1  &8.10 &0.07 &6.62 &0.04 &5.32 &0.04 &3.35 &0.05 &2.67 &0.04 \\
25.12.2025 & 19:10 &     &      &8.00 &0.06 &6.52 &0.05 &5.18 &0.02 &3.18 &0.04 &2.40 &0.05 \\
03.01.2026 & 14:40 &     &      &     &     &     &     &     &     &3.19 &0.02 &2.37 &0.03 \\
07.01.2026 & 14:20 &     &      &     &     &     &     &5.04 &0.10 &3.28 &0.04 &2.55 &0.06 \\
29.01.2026 & 16:00 &8.84 & 0.02 &7.73 &0.02 &6.39 &0.03 &5.21 &0.02 &3.27 &0.04 &2.52 &0.04 \\
08.03.2026 & 02:05 &     &      &     &     &     &     &     &     &3.35 &0.05 &2.56 &0.10 \\
\hline
 \end{tabular} \\
\end{table*}
% 		         %%%%%%%%%%%%%%%%%%%%%%%%%%%
%				%%%%%%%%%%%%%%%%%%%%%%%%%%%
\renewcommand{\thetable}{A3}
\begin{table*}
\renewcommand{\tabcolsep}{0.15cm}
 \caption{Polarimetric results for {\IRAS} in the $B,$ $V,$ $R_\mathrm{c},$ $I_\mathrm{c}$ bands.}
  \label{tab:polariz}
  \begin{footnotesize}
  \begin{tabular}{ll|llll|llll|llll|llll}
\hline
Date & UT & $p_B$ & $\sigma_p$ & $\theta_B$ & $\sigma_\theta$\ &
$p_V$ & $\sigma_p$ & $\theta_V$ & $\sigma_\theta$ &
$p_R$ & $\sigma_p$ & $\theta_R$ & $\sigma_\theta$ &
$p_I$ & $\sigma_p$ & $\theta_I$ & $\sigma_\theta$ \\
dd.mm.yyyy & hh:mm & \% & \% & $\degr$ & $\degr$ &
\% & \% & $\degr$ & $\degr$ & \% & \% & $\degr$ & $\degr$ \\
\hline
04.12.2025 &18:35 &     &     &     &     &      &    &     &    &      &     &     &    &13.85 &0.17 &65.3 &0.7 \\
09.12.2025 &16:17 &     &     &     &     & 7.26 &0.19 &65.0 &1.5 &8.10  &0.20 &64.0 &1.4 &13.96 &0.16 &65.2 &0.7 \\
18.12.2025 &18:04 &     &     &     &     & 8.74 &0.17 &66.1 &1.1 &9.42  &0.17 &65.6 &1.1 &15.02 &0.17 &66.4 &0.6 \\
21.12.2025 &16:30 &     &     &     &     & 9.25 &0.17 &66.1 &1.0 &9.97  &0.17 &65.6 &1.0 &15.68 &0.16 &66.5 &0.6 \\
03.01.2026 &14:57 &     &     &     &     & 9.41 &0.17 &70.2 &1.0 &10.39 &0.22 &68.2 &1.2 &15.86 &0.16 &68.2 &0.6 \\
07.01.2026 &15:06 &8.61 &0.23 &66.8 &1.6  & 9.67 &0.18 &68.4 &1.0 &10.66 &0.17 &68.3 &0.9 &15.53 &0.18 &66.6 &0.7 \\
\hline
 \end{tabular} \\
\end{footnotesize}
\end{table*}
% 		         %%%%%%%%%%%%%%%%%%%%%%%%%%%

\end{document}